\documentclass[aps,prb,twocolumn,groupedaddress,amsmath,amssymb]{revtex4}

\usepackage{amssymb}
\usepackage{amsmath}
\usepackage{graphicx}
\usepackage{subfigure}
\usepackage{textcomp}
\usepackage{color}
\usepackage{amsfonts}
\usepackage{bbold}
\usepackage{dsfont}
\usepackage{epsfig}
\usepackage{hyperref}

\newcommand{\arctanh}[1]{\operatorname{arctan}}

\begin{document}

\title{Polaronic distortion and vacancy-induced magnetism in MgO}

\author{A. Droghetti, C.D. Pemmaraju and S. Sanvito}
\affiliation{School of Physics and CRANN, Trinity College, Dublin 2, Ireland}

\date{\today}

\begin{abstract}
The electronic structure of the neutral and singly charged Mg vacancy in MgO is investigated using density functional theory. 
For both defects, semilocal exchange correlation functionals such as the local spin density approximation incorrectly predict a 
delocalized degenerate ground state. In contrast functionals that take strong correlation effects into account predict a localized 
solution, in agreement with spin resonance experiments. Our results, obtained with the HSE hybrid, atomic self-interaction corrected
and LDA+$U$ functionals, provide a number of constraints to the possibility of ferromagnetism in hole doped MgO.    
\end{abstract}

\maketitle

In recent years, the so called ``d$^0$ magnets'' have attracted great interest within the magnetism community and have 
been the topic of many theoretical and experimental investigations.  Broadly speaking, the term ``d$^0$ magnets'' refers 
to a class of materials which, lacking any magnetic ions with open $d$ or $f$ shells, should in principle not be ferromagnetic, 
but nevertheless exhibit signatures of ferromagnetism often with a Curie temperature exceeding 300 K \cite{1}. Defective 
graphite \cite{3} and gold nanoparticles capped with organic molecules \cite{4} are classical examples. However, the most 
widely studied class of d$^0$ magnets is that of un-doped or light-element doped medium to wide gap oxides. The first 
experimental study, which boosted research in this field, was on un-doped HfO$_2$ \cite{5}. Since then, there have been 
similar reports of ferromagnetism for numerous other oxides such as TiO$_2$~\cite{TiO2_5}, SnO$_2$~\cite{SnO2}, 
CeO$_2$~\cite{CeO2}, ZnO~\cite{ZnO}, CuO$_2$~\cite{CuO2}, MgO~\cite{MgO, MgO_1, MgO_2} etc.  Generally, the 
ferromagnetism is observed in highly defective samples leading to the expectation that the magnetism must be somehow 
defect-related. Nanostructures are also often reported ferromagnetic and it has been argued that ferromagnetism can be 
a general property of nanoparticles~\cite{nanop}.~Nevertheless the precise mechanism behind the reported high 
temperature ferromagnetism is not well understood and is a subject of open debate. 

On the theoretical front, \textit{ab initio} calculations based on Densisty Functional Theory (DFT) have been widely 
employed to study d$^0$ magnetism in a wide range of systems. Following the early work of Elfimov \textit{et al} on 
CaO~\cite{saw}, and in parallel with many experimental studies, a number of theoretical publications, proposing new 
d$^0$ ferromagnetic materials, have appeared over the past few years. In most of the d$^0$ oxides proposed, the 
formation of magnetic moments and the resulting ferromagnetism is attributed to spin-polarized holes residing on 
cation $p$-orbitals  either at vacany or impurity sites~\cite{saw,6}.  Critically however, almost all such predictions are 
based on DFT calculations performed with local approximations of the exchange correlation potential (LSDA or GGA). 
These notoriously suffer from the spurious self-interaction, whose consequence is that of over-delocalizing the charge 
density. It is then not surprising that most of the calculations return a metallic (often half-metallic) ground state and usually, 
extremely large magnetic interaction. It has been shown previously that correcting for self-interaction can lead to the 
holes states localizing as a consequence of lattice distortions and the magnetic coupling is then reduced drastically or even 
completely suppressed \cite{mine,me2}. This is consistent with the very well established notion that holes bound to acceptor 
defects form small polarons \cite{polaron1, polaron2}. More recently, the electronic properties of  cation vacancies in many Zn chalcogenides was investigated in detail by Chan \textit{et al}\cite{zunger2}. It was shown that the metallic band structure predicted 
by LDA is changed into an insulating one as soon as a hole-state potential operator \cite{zunger}, which increases the splitting 
between occupied and unoccupied anion p-states, is applied. Pardo \textit{et al}~\cite{Pickett} investigated the effect of on-site 
$U$ corrections on the $p$-orbital magnetism in N doped alkaline earth monoxides and showed that for realistic values 
of the $U$-parameter, the ferromagnetic coupling is drastically diminished. 

Without entering into the issues connected with percolation~\cite{zunger3}, already the fact that holes around acceptor 
defects form small bound polarons challenges the simple explanation of ferromagnetism due to carrier-mediated long-range 
coupling between magnetic moments residing on cation vacancies. The prototypical example of the cation vacancies in MgO 
clearly illustrates this point. Recently, numerous theoretical papers have proposed that Mg vacancies can induce room 
temperature ferromagnetism in MgO\cite{MgO_1, MgO_3, MgO_4}. Based mainly on DFT LSDA and GGA results, Mg 
vacancy centres are reported to produce half-metallic shallow acceptor states at the top of the valence band that lead to long
 range ferromagnetic coupling via a double exchange mechanism. 
 
However, the fact that a number of electron spin resonance (ESR) experiments dating back to the 1970s have clearly 
established the localized polaronic nature of Mg vacancies in MgO~\cite{Rose,Kappers} seems to have been 
forgotten. As pointed out by Stoneham \textit{et al}. \cite{polaron2}, the cation vacancy can be experimentally found in two
charging states: neutral, $\mathrm{V^0}_{\mathrm{Mg}}$, and singly charged, $\mathrm{V}^-_{\mathrm{Mg}}$. ESR 
measurements establish that $\mathrm{V}_{\mathrm{Mg}}$s, in both the charging states, are deep traps exhibiting states 
in the gap as opposed to shallow levels at the top of the valence band. In the case of $\mathrm{V}^0_{\mathrm{Mg}}$, the 
two holes from the vacancy localize completely on two adjacent oxygen sites. In contrast, in the case of 
$\mathrm{V}^-_{\mathrm{Mg}}$, the only hole is completely localized on one single oxygen anion with an accompanying 
bond distortion. 

The availability of ESR experimental data, establishing the nature of $\mathrm{V}_{\mathrm{Mg}}$ in MgO, sets a stringent test 
for DFT predictions obtained with different exchange correlation functionals. In particular the results of semilocal LSDA/GGA 
can be compared to beyond LDA approaches. In this work we look at three such methods. The first is the HSE hybrid-DFT 
functional~\cite{HSE}, which represents a new and promising tool to study many properties of solid state systems, in particular 
wide-gap oxides~\cite{kresse_hse}. This was recently shown able to describe correctly the interplay between electronic properties 
and lattice distortion in oxides as complex as BaBiO$_3$~\cite{BaBiO3}. The second is the atomic self-interaction correction (ASIC)
scheme, which has been employed previously to study a number of oxide and nitride wide-gap materials~\cite{ASIC}. The third is 
the well known LDA+$U$ method~\cite{Anisimov}.

LSDA, ASIC and LSDA+$U$ calculations were carried out using a development version of the SIESTA code~\cite{SIESTA}.  
Norm-conserving pseudopotentials with the following reference atomic configurations were employed. Mg([Ne]3s$^{2}$), 
([He]2s$^{2}$,2p$^{4}$). A double zeta basis set with additional polarization functions was used for both Mg and O. At the site of 
the vacany, extra basis functions corresponding to Mg atom were introduced to minimize basis set errors. 
Supercells containing 96 atoms and 192 atoms were used for single defect and multiple defect calculations respectively. Ionic 
coordinates were optimized using a conjugate gradients algorithm until all the forces are smaller than 0.04 eV/\AA.  
The GGA and HSE calculations were carried out using the VASP package~\cite{VASP}. A 54 atom supercell was used to 
study the  $\mathrm{V^0}_{\mathrm{Mg}}$ center, while a cubic 64 atom supercell was used to simulate the singly 
charged $\mathrm{V}^-_{\mathrm{Mg}}$ center. A planewave energy cutoff of 520 eV and a 3x3x3 $\Gamma$-centered 
$k$-point mesh were employed. For the HSE functional, the default value AEXX=0.25 is used for the Hartree-Fock exchange 
mixing parameter and for the screening of the Fock exchange HFSCREEN=0.2 is used. Geometry optimization was carried 
out using conjugate gradients.\\

\begin{table}[t]
\begin{tabular}{lcccc}\hline\hline
&\vline&$\mathrm{V^0}_{\mathrm{Mg}}$  &\vline&   $\mathrm{V}^-_{\mathrm{Mg}}$    \\ \hline
LSDA &\vline  & 2.19 (6) &\vline & 2.21 (6) \\    
GGA  &\vline& 2.233 (6) &\vline& 2.24 (6) \\
ASIC  &\vline& 2.11 (2), 2.22 (4)  &\vline& 2.13 (1), 2.24 (1), 2.25 (4) \\
HSE  &\vline&  2.111 (2), 2.253(4) &\vline& 2.106 (1), 2.225 (1), 2.253 (4) \\ \hline\hline
\end{tabular}
\caption{\label{Tab1} Calculated bond lengths (in \AA), for $\mathrm{V}^{0}_{\mathrm{Mg}}$ and 
$\mathrm{V}^{-}_{\mathrm{Mg}}$ as calculated by different DFT functionals. Bond lengths correspond 
to the distance from the center of the Mg vacancy site to the O ions in the co-ordination octahedron. 
The number in brackets is the number of bonds of a given length.}
\end{table}

\begin{figure}[ht]\centering
\includegraphics[scale=0.45,clip=true]{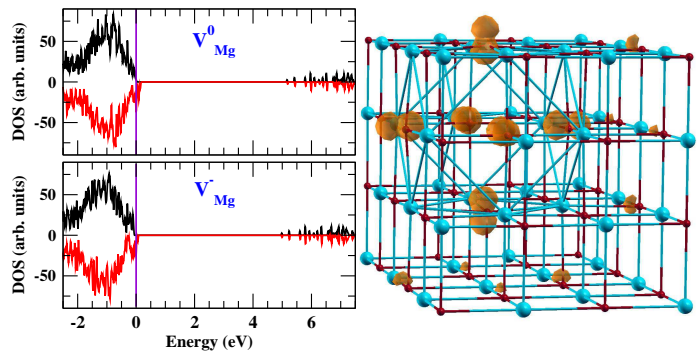}
\caption{LSDA electronic structure of $\mathrm{V_{MgO}}$. (Left) DOS of a MgO supercell containing either 
$\mathrm{V^0}_{\mathrm{Mg}}$ or $\mathrm{V}^-_{\mathrm{Mg}}$. (Right) Charge density corresponding to 
the minority spin holes at the top of the valence band for $\mathrm{V^0}_{\mathrm{Mg}}$. Note that the hole 
density is spread uniformly over the O octahedron.}
\label{Fig1}
\end{figure}
We start our analysis by presenting the LSDA results, with the GGA ones being rather similar. The $\mathrm{V}_{\mathrm{Mg}}$ site 
in MgO presents a perfect octahedral symmetry which is preserved in LSDA, even upon lattice relaxation. This is
the case for both $\mathrm{V^0}_{\mathrm{Mg}}$ and $\mathrm{V}^-_{\mathrm{Mg}}$. The bond lengths resulting from this 
isotropic relaxation are presented in table~\ref{Tab1}. The associated hole density always spreads uniformly over all of the six 
coordinating O anions with $p$-orbital lobes pointing towards the center of the vacancy (see Fig.~\ref{Fig1}). In the case of the 
$\mathrm{V^0}_{\mathrm{Mg}}$ a spin triplet, $S=1$, is obtained, while the spin singlet $S=0$ cannot be stabilized. In contrast
$\mathrm{V}^-_{\mathrm{Mg}}$ presents a $S=1/2$ ground state. For both the charging configurations the density of states (DOS) is 
half metallic as shown in Fig.~\ref{Fig1} and corresponds to that of a shallow acceptor impurity with hole states situated at the top 
of the valence band. This picture however does not conform with the experimental ESR data referenced earlier.  

\begin{figure}[ht!]\centering
\includegraphics[scale=0.5,clip=true]{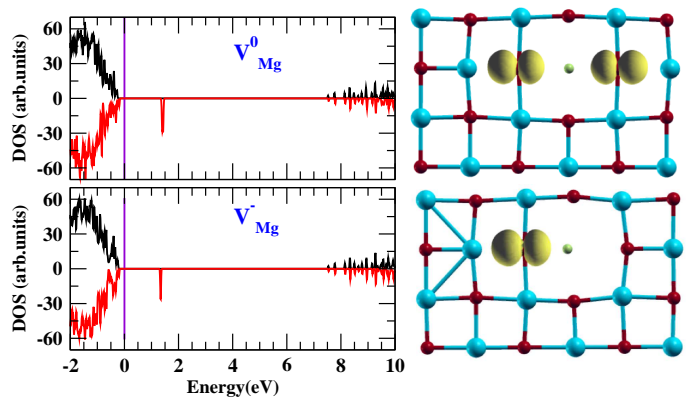}
\caption{ASIC electronic structure of $\mathrm{V_{MgO}}$. (Left) DOS of a MgO supercell containing either 
$\mathrm{V^0}_{\mathrm{Mg}}$ or $\mathrm{V}^-_{\mathrm{Mg}}$. (Top right) Charge density corresponding to the holes 
of $\mathrm{V^0}_{\mathrm{Mg}}$. Note that the two holes are completely localized on two opposing O ions. 
(Bottom right) Charge density corresponding to the single hole of $\mathrm{V^0}_{\mathrm{Mg}}$. The hole density 
is localized on one O ion.}
\label{Fig2}
\end{figure}
In contrast, a qualitatively different description of the ground state emerges from HSE and ASIC, with the two methods being
consistent with each other. Firstly, lattice relaxation around the vacancy site is seen to break the perfect octahedral symmetry. 
In the case of the  ($\mathrm{V^0}_{\mathrm{Mg}}$) center, two virtual bonds from the center of the vacancy to two opposing 
O anions are found to shorten with respect to four other bonds forming a plane perpendicular to the two shorter bonds. In the 
case of the $\mathrm{V}^-_{\mathrm{Mg}}$ site, one of the two previously shorter bonds is seen to elongate while the four fold 
symmetry of the bonds in the plane perpendicular to these bonds is retained (see table~\ref{Tab1}). 
The symmetry breaking around the vacany site is also accompanied by a strong localization of the hole density on the O ions 
nearest to the vacancy forming small polarons (see Fig.~\ref{Fig2}). It should be noted that the symmetry lowering around the 
vacancy site may not occur spontaneously during the course of the geometry optimization, since the octahedrally symmetric
solution is a local minimum. Therefore, in order to access the lower symmetry ground state it might be necessary to initialize 
the geometry optimization in a slightly symmetry distorted ionic configuration around the site. 

HSE returns us an energy gain due to the polaronic distortion of 360~meV for $\mathrm{V^0}_{\mathrm{Mg}}$ and 225 meV for 
$\mathrm{V}^-_{\mathrm{Mg}}$, and therefore it establishes that the distorted geometry is considerably more stable than the
octahedrally symmetric one regardless of the charging state. 
\begin{figure}[ht!]\centering
\includegraphics[scale=0.45,clip=true]{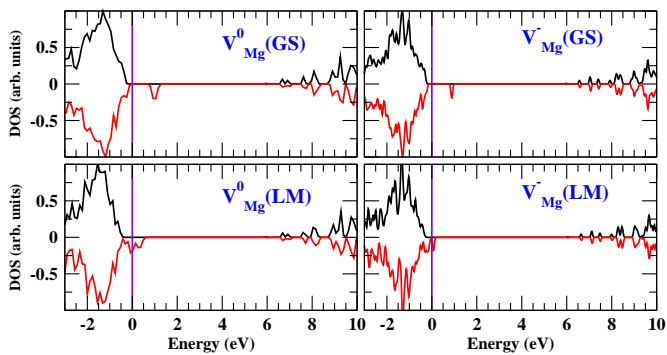}
\caption{DOS calculated with HSE for a MgO supercell containing either $\mathrm{V^0}_{\mathrm{Mg}}$ or 
$\mathrm{V}^-_{\mathrm{Mg}}$. The top two panels show the DOS for the polaronic ground state geometry 
(labelled GS) of the defects. The bottom two panels show the DOS corresponding to local minima solutions 
(labelled LM) which preserve the octahedral symmetry around the vacancy sites.}
\label{Fig3}
\end{figure}
The ground state DOS obtained within ASIC and HSE is shown in figures~\ref{Fig2} and ~\ref{Fig3}.  As opposed to the shallow 
acceptor states observed in LSDA/GGA, both  $\mathrm{V^0}_{\mathrm{Mg}}$ and  $\mathrm{V}^-_{\mathrm{Mg}}$ exhibit a deep 
acceptor level in the gap. This is located at $\sim$1~eV ($\sim$1.25~eV) above the valence band maximum (VBM) for HSE 
(ASIC). It is interesting to note that for the symmetrically relaxed local minimum solution within HSE (see lower panels of 
figure~\ref{Fig3}), a shallow acceptor DOS similar to the LSDA solution is obtained. Thus the deep acceptor ground state 
is intimately associated to the polaronic bond distortion around the vacancy.
\begin{figure}[ht!]\centering
\includegraphics[width=0.5\textwidth,clip=true]{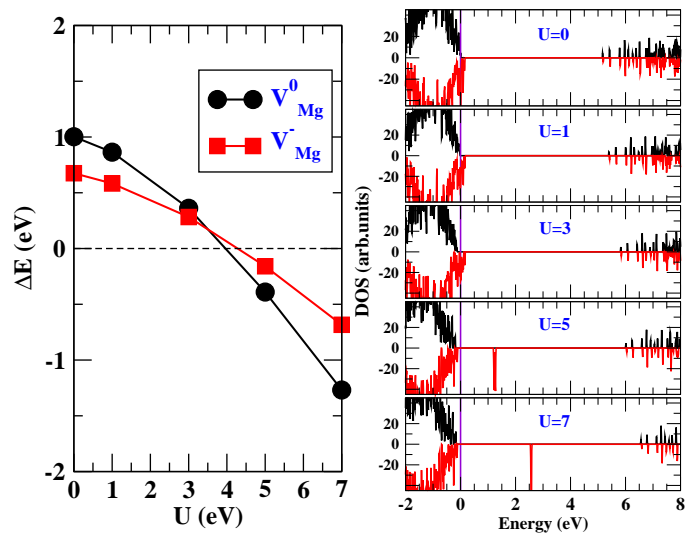}
\caption{(Left) LSDA+$U$ energy difference between the LSDA and ASIC geometries for 
$\mathrm{V^0}_{\mathrm{Mg}}$ and $\mathrm{V}^-_{\mathrm{Mg}}$ as a function of $U$. 
Note the crossover between the two geometries at $U\sim4$~eV, when the polaronic ASIC 
geometry becomes the most stable. (Right) Ground state DOS calculated with LSDA+$U$ as 
a function of $U$.}
\label{Fig4}
\end{figure}

In order to further substantiate the HSE/ASIC results we also carried out some test calculations within LSDA+$U$ mainly 
looking at the evolution of the ground state as a function of the strength of the $U$ parameter. 
The $U$ correction in this case is applied to the $p$-orbitals of the O sublattice. 
For a given value of $U$, the total LSDA+$U$ energies obtained at the LSDA and ASIC geometries presented in 
table~\ref{Tab1} are compared. The energy difference $\Delta E = E(\mathrm{G_{LSDA}})-E(\mathrm{G_{ASIC}})$, where 
$\mathrm{G_{LSDA}}$ ($\mathrm{G_{ASIC}}$) denote the LSDA (ASIC) geometry, is shown in figure~\ref{Fig4}. As the 
strength of $U$ is increased the bond distorted ASIC geometry is seen to be stabilized with respect to the symmetric LSDA 
geometry and the crossover is around $U=4$~eV for both  $\mathrm{V^0}_{\mathrm{Mg}}$ and $\mathrm{V}^-_{\mathrm{Mg}}$. 
Also the nature of the defect ground state changes from shallow acceptor to deep acceptor on either side of the 
crossover~(Fig.~\ref{Fig4}).  In wide-gap oxides, the value of $U$ for the O-2$p$ orbitals is estimated to be in the range of 
5-7~eV \cite{Pickett}. At $U$=5~eV, for the $\mathrm{V^0}_{\mathrm{Mg}}$ ($\mathrm{V}^-_{\mathrm{Mg}}$) center the polaronic 
$\mathrm{G_{ASIC}}$ geometry is lower in energy by a robust $\sim$390 meV (160 meV) with respect to the $\mathrm{G_{LSDA}}$ geometry. Furthermore the DOS for $\mathrm{V^0}_{\mathrm{Mg}}$~(Fig.~\ref{Fig4}) exhibits a deep acceptor state 
located at roughly 1.3~eV above the VBM which agrees well with the HSE results.       
   
We now look at the magnetic properties associated to Mg vacancies. For the $\mathrm{V^0}_{\mathrm{Mg}}$ center, 
the spins of the two holes may be aligned parallel or anti-parallel leading either to a $S=1$ triplet or a $S=0$ singlet. 
As discussed above, LSDA/GGA predict a stable $S=1$ configuration. However, within HSE/ASIC, the energies of the $S=1$ and 
the $S=0$ solutions are found to be very similar, separated only by 0.75 meV (2 meV) in HSE (ASIC).  We have also calculated the inter-defect magnetic 
coupling interaction between two ($S=1$) $\mathrm{V^0}_{\mathrm{Mg}}$ centers in a 192 atom supercell. For LSDA two 
sites separated by 8.34 \AA\ are ferromagnetically coupled with a spin-flip energy of 33~meV (the energy required to 
reverse the spin of one of the two vacancies).
This suggests the stabilization of ferromagnetism via magnetic percolation between the holes, provided they are 
formed at a sufficiently high concentration. In contrast ASIC predicts for the same supercell configuration, a tiny interaction 
energy of 1~meV. Thus, we find that as soon as the correct polaronic ground state is achieved for the holes, their magnetic 
interaction virtually disappears.

Before concluding, we add an additional remark on the similarity between $\mathrm{V^-}_{\mathrm{Mg}}$ and 
Li impurity substituting  Mg (Li$_{\mathrm{Mg}}$). Indeed, both  are single acceptors located at the cation sublattice. 
For Li$_{\mathrm{Mg}}$ as well, LSDA returns a half-metallic ground state, with the hole completely delocalized on 
all the O ions coordinating the impurity. ASIC instead localizes the hole on one O ion in a distorted polaronic geometry. 
This is qualitatively identical to what found for $\mathrm{V^-}_{\mathrm{Mg}}$ (the Li-O bond lengths are: 2.31 \AA (1), 
1.93~\AA (1), 2.15~\AA (4)). Although we did not find an experimental reference to confirm the ground state of 
Li$_{\mathrm{Mg}}$ in bulk MgO, its nature as a small bound polaron is well established in the case of surfaces and 
beyond-LSDA electronic structure methods have previously been used to study this defect in the context of 
catalysis~\cite{watson}.

In conclusion, we have shown that semilocal LSDA/GGA can lead to qualitative failures in their description of 
hole centers in MgO. Beyond-LDA approaches that are either self interaction free or effectively correct for it, 
are able to reproduce the experimentally observed polaronic ground state. Significantly, the large inter-site 
ferromagnetic interaction predicted by LSDA is also shown to be an artifact. Thus, the observed high temperature 
ferromagnetic signals in MgO cannot be explained by a simple model where magnetic interactions between 
hole centers, in their ground state, percolate through the sample.    

This work is supported by Science Foundation of Ireland and by the EU (ATHENA project). Computational resources
have been provided by the Trinity Center for High Performance Computing and by ICHEC.


\begin{thebibliography}{100}
\bibitem{1} J.M.D.~Coey, Solid State Sci., \textbf{7}, 660 (2005). 

\bibitem{3}T.~Makarova  and F.~Palacio (ed) {\it Carbon-Based Magnetism: An Overview of the Magnetism of Metal-Free Carbon-Based Compounds and Materials} (Elsevier, Amsterdam, 2006).


\bibitem{4}P.~Crespo, R.~Litran, T.C.~Rojas, M.~Multigner, J.M.~de~la~Fuente, J.C.~Sanchez-Lopez, M.~A.~Garcia, A.~Hernando, S.~Penades, A~Fernandez, Phys. Rev. Lett., \textbf{93}, 087204 (2004).

\bibitem{5} M.Venkatesan, C.B. Fitzgerald, J.M.D. Coey, Nature, \textbf{430}, 630 (2004).


\bibitem{TiO2_5} S.~Zhou, E. Cizmar, K. Potzger, M.~Krause, G.~Talut, M.~Helm, J.~Fassbender, J. Wosnitza, H.~Schmidt, Phys. Rev. B, {\bf 79}, 113201 (2009).

\bibitem{SnO2} R.~P.~Panguluri, P.~Kharel, C.~Sudakar, R.~Naik, R.~Suryanarayanan, V.~M~Naik, A.~G.~Petukhov, B.~Nadgorny and G.~Lawes, Phys. Rev. B {\bf 79}, 165208 (2009).


\bibitem{CeO2} Y.~Liu, Z.~Lockman, A.~Aziz, J. MacManus-Driscoll, J. Phys.: Condens. Matter, {\bf 20}, 165201 (2008). 

\bibitem{ZnO} K.~Potzger, S.~Zhou, J.~Grenzer, M.~Helm, J.~Fassbender, Appl. Phys. Lett., {\bf 92}, 182504 (2008).


\bibitem{CuO2} C.~Chen, L.~He, L.~Lai, H.~Zhang, J.~Lu, L.~Guo, Y.~Li, J. Phys.: Condens. Matter, {\bf 21}, 145601 (2009).

\bibitem{MgO} J.~Hu,Z.~Zhang, M.~Zhao, H.~Qin, M.~Jiang, Appl. Phys. Lett. {\bf 93}, 192503 (2008).

\bibitem{MgO_1}J.~I.~Beltr\'{a}n, C. Monty, Ll. Balcells , C. Mart\'{\i}nez-Boubeta, Solid State Comm. \textbf{149}, 1654 (2009).

\bibitem{MgO_2}N.~Kumar, D. Sanyal, A. Sundaresan, Chem. Phys. Lett. \textbf{477}, 360 (2009).


\bibitem{MgO_3}F.~Gao, J.~Hu, C.~Yang, Y.~Zheng, H.~Qin, L.~Sun, X.~Kong and M.~Jiang, Solid State Comm. \textbf{149}, 855 (2009).


\bibitem{MgO_4}F.~Wang, Z.~Pang, L.~Lin, S.~Fang, Y.~Dai and S.~Han, Phys. Rev. B \textbf{80}, 144424 (2009).

\bibitem{nanop} A.~Sundaresan, R.~Bhargavi, N.~Rangarajan, U.~Siddesh and C.~N.~R.~Rao, Phys. Rev. B {\bf 74}, 161306(R) (2006).

\bibitem{saw} I.S. Elfimov, S.Yunoki, G.A. Sawatzky Phys. Rev. Lett. \textbf{89}, 216403 (2002).

\bibitem{6}C.~D.~Pemmaraju and S.~Sanvito, Phys. Rev. Lett {\bf 94}, 217205 (2005).

\bibitem{mine}A.~Droghetti, C.~D.~Pemmaraju and S.~Sanvito, Phys. Rev. B, {\bf 78}, 140404(R) (2008).

\bibitem{me2} A.~Droghetti and S.~Sanvito, Appl. Phys. Lett. {\bf 94}, 252505 (2009). 

\bibitem{polaron1}O.F.~Schirmer, J. Phys.: Condens. Matter, {\bf 18}, R667 (2006).


\bibitem{polaron2} A.M.~Stoneham, J. Gavartin, A.L.~Shluger, A.V.~Kimmel, D.~Mun\~oz~Ramo, H.M.~R\o nnow, G.~Aeppli, C.~Renner, J. Phys.: Condens. Matter, {\bf 19}, 255208 (2007).

\bibitem{zunger2}J.A.~Chan, S.~Lany, A.~Zunger, Phys. Rev. Lett. {\bf 103}, 016404, (2009).

\bibitem{zunger}S.~Lany and A.~Zunger, Phys. Rev. B {\bf 80}, 085202, (2009).

\bibitem{Pickett}V. Pardo and W. E. Pickett, Phys. Rev. B \textbf{78}, 134427 (2008).

\bibitem{zunger3}J.~Osorio-Guill\'en, S.~Lany, S.V.~Barabash, A.~Zunger, Phys. Rev. Lett., {\bf96}, 107203 (2006).

\bibitem{Rose}B.~H.~Rose and L.~E.~Halliburton, J. Phys. C: Solid State Phys. \textbf{7}, 3981 (1974).

\bibitem{Kappers}L.~A.~Kappers, F.~Dravnieks and J.~E.~Wertz, J. Phys. C: Solid State Phys. \textbf{7}, 1387 (1974).

\bibitem{HSE}J.~Heyd, G.~E.~Scuseria, and M.~Ernzerhof, J. Chem. Phys. \textbf{118}, 8207 (2003).


\bibitem{kresse_hse}J.~Paier, M.~Marsman, K.~Hummer, G.~Kresse, I.C.~Gerber and J.G.~\'Angy\'an, J.Chem. Phys. \textbf{124},154709 (2006).

\bibitem{BaBiO3} C.~Franchini, G.~Kresse and R. Podloucky, Phys. Rev. Lett.  \textbf{102}, 256402 (2009).

\bibitem{ASIC}C.~D.~Pemmaraju, T.~Archer, D.~S\'{a}nchez-Portal, and S.~Sanvito, Phys. Rev. B \textbf{75}, 045101 (2007).

\bibitem{Anisimov}V.~I.~Anisimov, F.~Aryasetiawan, and A. I. Lichtenstein, J. Phys.: Condens. Matter \textbf{9}, 767 (1997).


\bibitem{SIESTA}J. M. Soler, E. Artacho, J. D. Gale, A.~Garc\'{\i}a, J.~Junquera, P.~Ordej\'{o}n and D.~S\'{a}nchez-Portal, J. Phys.: Condens. Matter \textbf{14}, 2745 (2002).

\bibitem{VASP}G. Kresse and J. Hafner, Phys. Rev. B \textbf{48}, 13115 (1993).

\bibitem{watson}M.~Nolan, G.~Watson, Surf. Science, {\bf 586}, 25 (2005).
\end{thebibliography}
\end{document}